\documentclass[letterpaper,twocolumn,english,showpacs]{revtex4}
\usepackage[]{fontenc}
\usepackage[latin1]{inputenc}
\usepackage{graphicx}

\makeatletter

\usepackage{graphicx}

\usepackage{graphicx}

\usepackage{graphicx}

\usepackage{graphicx}

\usepackage{babel}
\makeatother
\begin{document}

\title{Spectroscopy of phonons and spin torques in magnetic point contacts}

\author{I. K. Yanson, Yu. G. Naidyuk, D. L. Bashlakov, V. V. Fisun, O. P.
Balkashin}

\address{B. Verkin Institute for Low Temperature Physics and Engineering,
National Academy of Sciences of Ukraine, 61103, Kharkiv, Lenin av.
47, Ukraine}

\author{V. Korenivski}

\address{Nanostructure Physics, Royal Institute of Technology, 10691 Stockholm,
Sweden}

\author{R. I. Shekhter}

\address{Department of Applied Physics, Chalmers University of Technology,
Göteborg, Sweden}

\begin{abstract}
Phonon spectroscopy is used to investigate the mechanism of current-induced
spin torques in nonmagnetic/ferromagnetic (N/F) point contacts. Magnetization
excitations observed in the magneto-conductance of the point contacts
are pronounced for diffusive and thermal contacts, where the electrons
experience significant scattering in the contact region. We find no
magnetic excitations in highly ballistic contacts. Our results show
that impurity scattering at the N/F interface is the origin of the
new single-interface spin torque effect.
\end{abstract}


\maketitle
Electrical point contacts are known to be an efficient tool for studying
electronic scattering in conductors, taking place on the nanometer
length scale \cite{PCSbook}. High electrical current concentrations
that can be obtained in nanoconstrictions allow to discriminate various
electronic relaxation mechanisms in the system. Three regimes of electron
current flow through a point contact are distinguished, depending
on the size of the contact in relation to the characteristic scattering
lengths. A ballistic regime occurs for very small contacts, with the
size smaller than any scattering length in the material. In this case
the resistance of the system does not depend on any material specific
dissipation and is determined purely by the geometry (the so-called
Sharvin resistance \cite{sharvin}). A diffusive regime corresponds
to a significant elastic scattering of electrons on impurities in
the point contact area, with the inelastic electron scattering length
still exceeding the size of the contact. Finally, a thermal regime
occurs for relatively large contacts where both elastic and inelastic
scattering of electrons take place within the contact, leading to
a significant local heating.

Electron transport in magnetic point contacts has become a focus of
intense theoretical and experimental study after the seminal predictions
that the magnetization of an inhomogeneous ferromagnet can be strongly
affected by a current of high density \cite{Slon,Berger}. The electron
spin plays the role of a mediator, transferring magnetization between
non-collinear ferromagnetic regions, which can lead to current-driven
magnetization precession as well as switching. These magnetic excitations
in turn affect the transport through the Giant Magneto-Resistance
effect \cite{baibich}. Current-induced magnetization precession and
switching have been observed in a number of experiments on magnetic
multi-layers \cite{Tsoi,TsoiN,Myers,Katine,kiselev,Rippard}. It has
recently been proposed that such current induced magnetization excitations
should occur even for single ferromagnetic layers \cite{Polianski,stiles}.
The new mechanism, in contrast to that of \cite{Slon,Berger}, relies
on spin transfer in the direction normal to the current flow. The
spin transfer is mediated by electrons, which are spin-dependently
reflected from the N/F interface, as illustrated in Fig. 1(a). The
spin-dependent reflection can be viewed as a transfer of a magnetic
moment $\delta m_{1}$ from the ferromagnet to the backscattered electron,
which becomes spin-polarized. Impurities in N scatter this electron
back at the interface. The second incidence at the N/F interface results
in a spin transfer between point 1 and point 2 of the interface (see
Fig.\,1(a)), such that $\delta m_{1}\neq\delta m_{2}$. It is predicted
that such transverse spin transfer mediated by backscattered electrons
at the two interfaces of thin ferromagnetic layers having asymmetric
normal metal electrodes can result in a net spin torque, sufficient
for exciting spin waves in the direction normal to the current flow
\cite{Polianski}. It is also predicted that, even in the absence
of such two-interface asymmetry, magnetic excitations can occur if
the spin distribution in F at the N/F interface is non-uniform \cite{stiles}.
Such single-layer, or single-interface, spin torques have been used
to interpret singularities in the differential resistance of N/F point
contacts and nano-pillars reported recently \cite{Ji,Oezyil}.

\begin{figure}
\includegraphics[%
  width=0.9\columnwidth]{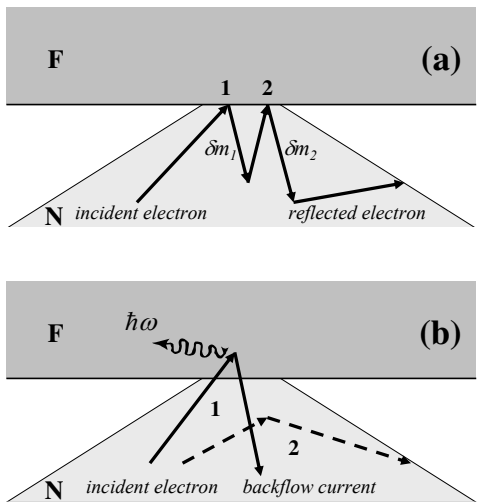}

\caption{Schematic of: (a) the single-interface spin transfer mechanism (here
between points 1 and 2 of F); (b) inelastic phonon scattering in ballistic
(trajectory 1) and diffusive (trajectory 2) point contacts.}
\end{figure}

The key feature in the above single-interface effect is the electron
backscattering on impurities near the N/F interface. Another phenomenon
originating from the backscattering of electrons in point contacts
is inelastic relaxation, which constitutes the basis of Point Contact
Spectroscopy (PCS) \cite{PCSbook}. PCS resolves relatively rare inelastic
scattering events for electrons having excess energy controlled directly
by the voltage applied to the contact \cite{kulik}. Such scattering
events induce a back-flow electron current, represented by trajectory
1 in Fig. 1(b). The second derivative of I(V) - the so-called point
contact spectrum - reflects the energy dependent coupling of the conduction
electrons with the elementary excitations in the material. Interactions
of electrons with phonons, magnons \cite{kulik1}, and other excitations
in metals, semiconductors, superconductors, and magnets have been
successfully studied using PCS \cite{PCSbook}. In contrast to the
single-interface spin torque effect, impurity scattering of electrons
affects PC spectra destructively, diminishing the back-flow of phonon-scattered
electrons (see trajectory 2 in Fig. 1(b)). The result is less pronounced
PC spectra (phonon peaks) as the mean free path is reduced and becomes
smaller than the size of the contact. Thus, PCS is a direct probe
of the intensity of impurity scattering in a point contact.

A direct experimental verification of the new mechanism of spin torque,
induced by backscattered electrons, would probe independently the
strength of impurity scattering at the N/F interface and correlate
it with the strength of the magnetic excitations. In this letter we
present such a study using PCS, correlating the regime of current
flow with the magneto-conductance. Our results clearly show that the
current-induced magnetization excitations are pronounced in the diffusive
and thermal regimes and practically absent in the ballistic regime.
This observation is a direct evidence of the impurity origin of the
single-interface spin torque effect, and serves as a clear demonstration
of the theoretical predictions \cite{Polianski,stiles}.

The current-voltage characteristic (IVC), $I(V)$, and its first,
$dV/dI(V)$, and second, $d^{2}V/dI^{2}(V)$, derivatives were measured
for hetero-contacts between needle-shaped non-magnetic Cu and Ag and
ferromagnetic Co in both bulk and film form. The film samples were
100 nm Co layers on 100 nm of Cu, acting as the bottom electrode,
both e-beam evaporated onto oxidized Si substrates. Most contacts
were made using a sharpened 0.15 mm diameter Ag wire, with the micro-positioning
and contact pressure controlled from outside the cryostat. All measurements
were done at 4.2 K, with the contacts created and measured directly
in liquid helium. We have recorded hundreds of PC spectra, which are
illustrated and analyzed below.

\begin{figure}
\includegraphics[%
  width=1.0\columnwidth]{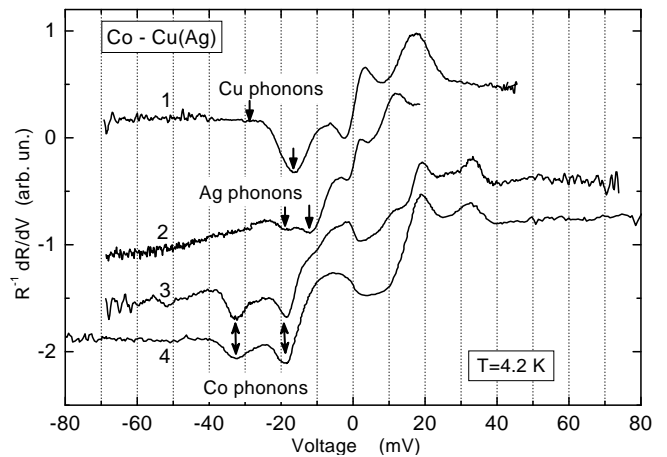}

\caption{PC spectra of hetero-contacts between bulk Co and Cu (curve 1, $R=31.5\,\,\Omega$)
or Ag (curve 2 and 3, $R=5.7\,\,\Omega$ and $R=8.1\,\,\Omega$) in
zero magnetic field. Curve 4 is the spectrum of a contact with $R$=5.2$\Omega$
between a 100 nm Co film and a Cu needle. The modulation voltage is
$V_{1}\approx1$ mV rms. The curves are offset vertically with respect
to curve 1 for clarity. The arrows indicate the positions of the main
phonon peaks of the corresponding metals (see \cite{PCSbook}). The
spectra are independent of field.}
\end{figure}

Fig. 2 shows PC spectra, $R^{-1}dR/dV\propto d^{2}V/dI^{2}$, of Co/Cu(Ag)
hetero-contacts, which, according to theory \cite{shekhter}, represent
a sum of partial contribution of both metals. The spectra have the
usual form of symmetric about $V=0$ extrema, observed at voltages
corresponding to the maxima of the electron phonon interaction (EPI)
function in Co and Cu or Ag \cite{PCSbook}. The intensity of the
phonon peaks varies depending on the exact position of the N/F boundary
with respect to the narrowest point of the nano-constriction (typically
$\sim10$ nm %
\footnote{Using the Wexler formula (see, e. g., \cite{PCSbook} Eq. 3.18), the
electron density in Co from \cite{gribov}, and the Co film resistivity
of 5 $\mu\Omega$cm, the PC diameter is estimated to lie between 10
and 15 nm for the contact resistance between 10 and 5 $\Omega$, respectively.%
}) as well as on the strength of EPI of the interfaced metals. The
spectra showed predominantly Co peaks located at approximately 19
and 33 meV, in agreement with earlier point contact investigations
\cite{gribov}. This indicates that the N/F interface was within the
point contact since the strength of EPI in Co is several times higher
than that in the noble metals \cite{gribov}, the effect favoring
the Co spectral contribution in a symmetric contact. Infrequent spectra,
dominated by Cu or Ag phonon peaks, are illustrated by curves 1 and
2 in Fig. 2 and indicate a deeper penetration of the noble metal in
to the Co electrode. Curve 3 illustrates the intermediate case, where
the phonon features of both metals are resolved. The S and N-shaped
features near zero bias are typical for scattering on magnetic impurities
(likely Co atoms in N - the so-called Kondo anomaly \cite{PCSbook}).
Thus, the spectra provide detailed information on the composition
and purity of the contacts. Same spectral features were observed for
point contacts to Co films, as illustrated by curve 4 in Fig. 2.

Fig. 3 shows a PC spectrum for a point contact between a bulk Co sample
and a Cu needle. The electron current flow is non-ballistic as witnessed
by the smeared out kink at positive bias of about 16 mV, which is
characteristic for Cu phonons \cite{PCSbook} in a non-spectroscopic
regime (close to the thermal regime). Two maxima in $dV/dI$, clearly
resolved as N-shaped features in the second derivative, are observed
on the negative bias branch, corresponding to electrons flowing into
the ferromagnet. These maxima are similar to those reported recently
\cite{Ji} and correspond to steps in the static resistance of the
contact of typically several percent. Their width in bias voltage
is smaller than the spectral resolution for the modulation used, indicating
their threshold rather than spectral character. They are shifted to
high bias by magnetic field, as shown in the inset to Fig. 3, with
a slope of 0.3--1 mA/T typical for other spin torque studies \cite{Rippard,Ji}.
Field sweeps at fixed near-threshold bias produce characteristic bell-shaped
magnetoresistance (not shown). All these features point to magnetic
excitations as the origin for the observed effect. Similar to \cite{Ji,stiles}
we assume a non-uniform spin distribution in F near the point contact,
induced or enhanced by the high density current through the contact,
which in turn gives rise to magnetoresistance. The characteristic
length scales in Co (the exchange length, spin diffusion length, domain
wall thickness) are all shorter than 100 nm, so spin profiles in the
point contact region ($\sim10$ nm in size) for bulk Co and 100 nm
Co films are expected to be essentially the same.

\begin{figure}
\includegraphics[%
  width=1.0\columnwidth]{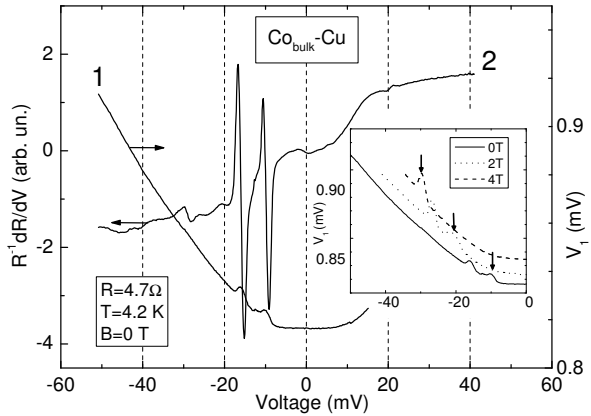}

\caption{First (1) and second (2) derivatives of the IVC for a hetero-contact
of bulk Co-Cu, $R$=4.7 $\Omega$. Inset shows the dependence of the
first maximum (shown by arrows) in modulation voltage $V_{1}\propto dV/dI$
on external magnetic field.}
\end{figure}

It is interesting to compare point contacts having the same resistance,
and therefore same current density in the nano-constriction, but different
electron flow regimes. Such a comparison is shown in Fig. 4 for three
contacts to both bulk and film Co. The N-shaped magnetic peaks are
pronounced for the diffusive contacts (having weak phonon features
and high background levels, solid lines) and practically absent for
the ballistic contacts (having well resolved phonon peaks and relatively
low background levels, dashed lines). These data clearly show that,
for a given current density, the probability of observing current-induced
magnetic excitations is determined by the strength of the impurity
scattering in the contact near the N/F interface. This observation
is qualitatively the same for bulk Co and Co films, which indicates
that the observed behavior is a single-interface effect.

\begin{figure}
\includegraphics[%
  width=1.0\columnwidth]{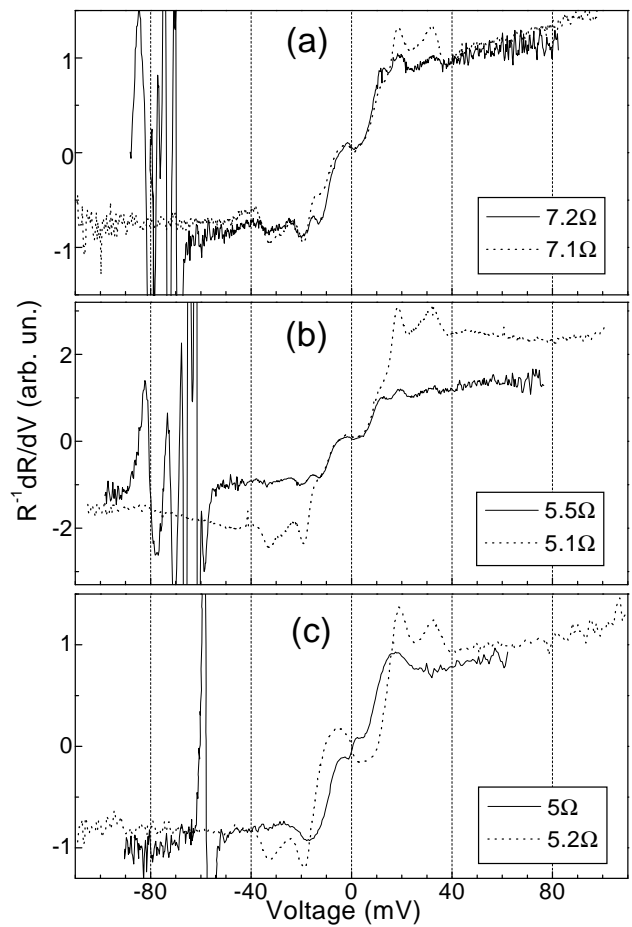}

\caption{A comparison of the PC spectra for pairs of contacts of nearly equal
resistance but different current flow regimes: (a) and (b) - spectra
between bulk Co and Ag taken in zero field; (c) - spectra between
100 nm Co film and Ag (solid line, B=3T) or Cu (dotted line, B=4T).
The threshold current density for the magnetic peaks is $5-6\times10^{9}$
A/cm$^{2}$.}
\end{figure}

The above conclusion on the role of impurity scattering in the single-interface
spin torque effect is supported by our measurements on a large number
of point contacts. The statistics of observing magnetic excitations
is shown in Fig. 5 as a function of the spectral quality of the contacts.
The latter is characterized by a phenomenological $\gamma$ parameter,
the inverse of which is the strength of the phonon peaks relative
to the background (see the inset to Fig. 5). The probability of observing
the current induced magnetic excitations for ballistic ($\gamma<1$)
versus non-ballistic (diffusive or thermal, $\gamma>1$) contacts
clearly demonstrates that the ballistic regime disfavors spin transfer
torque effects (open squares). Note, that a diffusive character of
current flow through a contact may not be a sufficient condition for
occurrence of magnetic instabilities. The relatively rare diffusive
contacts without magnetic peaks (4/17$\approx$24\% for $\gamma>1$)
may result from some unfavorable spin distribution in the ferromagnet
(a non-uniform, domain like spin distribution is required for observing
magneto-conductance \cite{Ji}). On the other hand, among hundreds
of point contacts measured, we have never observed a single highly
ballistic contact that would exhibit magnetic excitations (0 of 6
contacts with $\gamma<0.8$). Thus, our results provide a direct evidence
for the recently proposed single-interface mechanism of current-induced
spin torques \cite{Polianski,stiles}, which relies on impurity mediated
'secondary' polarization of a nominally \emph{un-polarized} current
incident on a N/F interface and should be absent in the ballistic
regime.

\begin{figure}
\includegraphics[%
  width=1.0\columnwidth]{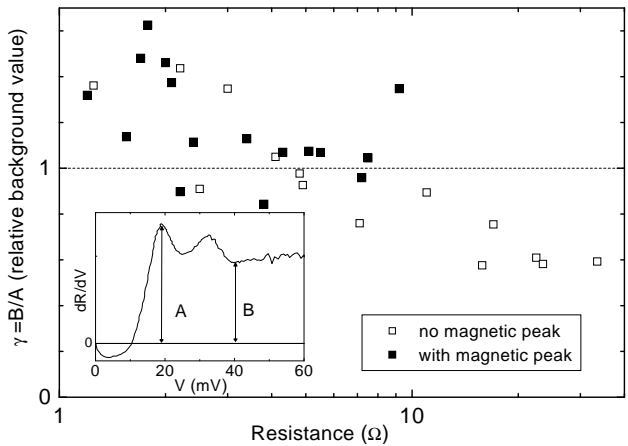}

\caption{Background value $\gamma$ for point contact spectra with magnetic
peak(s) (solid squares) and without magnetic peak(s) (open squares).
Inset: $\gamma$ is the ratio of the background intensity at around
40 mV to the intensity of the main phonon maximum at around 20 mV
(Co).}
\end{figure}

In summary, we have used phonon spectroscopy to study the mechanism
of the single-interface spin torque effect in non-magnetic/ferromagnetic
point contacts. In contrast to spin torques in magnetic multilayers,
this new mechanism relies on strong impurity scattering near the N/F
interface. We find no magnetic excitations in highly ballistic contacts
and pronounced spin torque effects in contacts where the electron
flow is diffusive or thermal. Our results provide a direct evidence
for the mechanism of the single-interface spin torque effect and should
be useful for design of spin torque devices based on single magnetic
layers or particles.

\begin{acknowledgments}
Financial support from the Swedish Foundation for Strategic Research
(SSF), the Royal Swedish Academy of Sciences (KVA), and the National
Academy of Sciences of Ukraine (NASU) under project \char`\"{}Nano\char`\"{}
10/05-H are gratefully acknowledged. The authors would like to thank
L. Gorelik and J. Slonczewski for fruitful discussions.
\end{acknowledgments}

\end{document}